**Assessing and Supplying the Health of Videos Games via Formal Semantics**

Mohammad Reza Besharati[1], PhD Candidate, besharati@ce.sharif.edu

Mohammad Izadi[1], Associate Professor

1- Department of Computer Engineering, Sharif University of Technology, Tehran, Iran

**Abstract**

Video games, just like any other media have both explicit and implicit messages, and they can have impact on physical and mental health of the users. These impacts can be positive or negative. The impacts, the implications and the meanings which exist in a game can be very widespread, multilayered and complicated. To investigate and guarantee the health of these video games, it is necessary to be able to estimate, assess and determine the implications of video games (from different perspectives). A common approach for studying complicated and multilayered phenomenon is formal semantics. Formal and rigorous methods can help in assessment and supplying the health of video games. In this article, an organizing for this assessment is proposed which is based on formal and rigorous methods and it considers various beneficiaries' concerns. Moreover, a technological solution is presented which is based on system compliance to meanings, model checking methods and logical solution. The proposed organizing has several features such as: agility, flexibility, scalability, repeatability of reviews, transparency, adaptation, available details for reviews, assessing various layers and implicit and explicit implications of system of the game, avoiding subjectivity or individual skills, relying on rules and regulations, ability to plan for beneficiaries because of its transparency and employment for specialists.

1. **Introduction**

Today, video games are not just simple games designed for pure entertainment; instead gaming industry which is a complicated and profitable industry is considered as one of the most important contexts for economy, knowledge and society. Game is cultural product and more important, it is a human need for one of the most sensitive layers of society (i.e. children, teenagers and youngsters). Gaming is so connected to all dimensions of social and individual life of a person that there is no similar example for this connection. Of course, we mean video games which, in less than half a century, became one of the most complicated feature influencing the lives of humans now and in future.

### 1.1. Aspects of the Impacts of Video Games and the Necessity of their Health

Video games, just like any other media have both explicit and implicit messages [1]. The impacts, the implications and the meanings which exist in a game can be very widespread, multilayered and complicated. In this section of the article, we deal with those aspects of video games which are related to the impacts on the player. The presented aspects are just some of various aspects of the impacts of video games, and they are presented only as samples to emphasize on the importance of the health of these games.

Excitement and being exciting is one of the most important factors in a game desirability [2]. Even in some research, excitement is considered as the most important factor in choosing a game [2]. Mental excitements (like those exist in fighting games) influence other mental activities and they get the mind involved long after the game is finished. Then, they disrupt other activities [2].

Excitement and violence are two different factors. However, many games provide excitement through anger for the player [3]. An important necessity for games is their limited violence.

Excitement has different forms and each form has different impact on mind and body. For example, the impact of excitement on brain, immune system and endocrine system [3] or the excitements which are mixed with stress can lead to a defect in memory [3]. Experiencing extreme excitement can disturb cognitive function, decrease the quality of decision making and disrupt an individual focus [3]. Stressful excitements make a person organize information incompletely, or not allocating enough time to investigate other solutions; in this condition, the person will decide before investigating all possible solutions [3]. In Islamic Ethics perspective, extreme excitement can decrease deliberation, meekness and prudence.

Extreme excitement which are produced by continuous stress can cause dendritic degeneration of some areas of hippocampus via hormonal mechanisms; this will suppress the neurogenesis of dentate gyrus neurons in this area [3]. (i.e. They harm the brain physiologically). The existing excitements in video games can harm immune system with affecting endocrine system [3]. These kinds of excitements are a lot in today video games, and they should be taken care of, especially when it comes to children and teenagers. Computers and video games can impose extreme excitements to their users and in this way, they can be dangerous for the health of the users [3].

Designing games with a focus on educational and scientific aspects can be useful in strengthening mental and cognitive skills, and they can be used as an educational tool [2]. For example, psychologists believe that immediate feedback and emotional control which are experienced by a player in an adventure game increase self-esteem and make the players believe in their abilities and qualifications

[2]. Some research show that strong, intermediate and weak students choose different kinds of games [2]. For instance, they show that strong students do adventure games more than other games, and this is related to the fact that these games are creativity-based [2]. These samples confirm the fact that video games can have positive and negative impacts on mind and body of players, and managing these impacts properly is a very important task. Therefore, identifying, defining, supplying and guarantee the health of video games is both important and desirable.

Health management in gaming industry and market, like other industries and markets, is possible only if all involved parties (including society and its other institutions such as religious and cultural institutions, government, media, regulatory authorities, game developers and gaming industry, users and parents) cooperate and integrate.

### 1.2. Semantics for Investigating Video Games

As it was mentioned before, video games have both implicit and explicit messages [1]. In semantics, implicit messages are those meaningful structures which are highly abstract and combine more tangible meanings. The impacts, the implications and the meanings which exist in a game can be very widespread, multilayered and complicated. To investigate and guarantee the health of these video games, it is necessary to be able to estimate, assess and determine the implications of video games (from different perspectives).

From semantic point of view, a game can be considered as its representation. Observing each representation of each game, we have different aspects of its implications. For example, class diagrams for a video game provide implications of source codes of the game, and game scripts provide implications of the narrative and story of the game.

Representation can be categorized in three levels: description, specification and specialization. Description is a form of representation which has enough flexibility and management for the abstract complications of a game and, it also has some extent of accuracy and formal implication. In this categorization, logic and logical representation is a description. So it is common to exchange logical representations which logical or formal descriptions.

Four levels can be identified for semantics: case, systematic, rigorous and formal. Formal description is based on a formal semantics while a systematic description is based on a systematic semantics. For example, UML diagrams can be considered as systematics descriptions, or Z Notation descriptions are based on Z formal semantics.

Till now, games have been more studied semantically as a text [1], narrative and story [1], art work, software, media, cognitive interactive system, conceptual system and other things like this. They have been less studied, modeled and analyzed as a logical and computational system. As a result, the level of semantic reviews of games has not exceed from systematic level; and we could not review games formally or rigorously.

To provide regulation for the health in game industry, and study the compliance of the games to this regulations, it is necessary to be able to study, model and analyze video games rigorously or formally. Studying compliance is an important task (it has legal, financial and other dimensions in industry), to supply this, we need rigorous and formal semantics and systematic semantics is not enough in this field. Subjective, vague or wrong encounter is more probable in systematic semantics than rigorous and formal semantics. Rigorous and formal semantics allow automation of compliance investigation.

**Table (1): Representation and Semantic Form for Some of the Common Artifacts in Game Industry**

| Artifacts | Representation Form | Semantic From |
|---|---|---|
| Game Script | Description | Case or Systematic |
| Class Diagram | Specification | Systematic |
| Game Characters' Graphic Design | Specialization | Rigorous |
| Game Code | Specialization | Rigorous |
| Logical and Computational Description | Description | Rigorous or Formal |

Compliance investigation manually for complicated logical and computational systems is very hard and even impossible, so its automation is very important. Games are also complicated soft systems in which compliance investigation automation is very important, and this fact can be achieved through rigorous and formal semantics.

The demands, requirements and features of games – considering every aspects of them such as technical, cognitive, artistic and so on _ is becoming more and more complicated. Using rigorous and formal methods can help manage these different demands wisely and consciously; they can decrease the use of methods which are solely based on try and error. Moreover, they provide higher levels of automation and quality guarantee for a game industry with these complicated products.

Now it is time for Serious Games, Games with a Purpose (GWAP), Educational Games (a part of edutainment discourse), VR Games and Cognitive Games (which are used to improve cognitive and problem solving skills). This fact makes studying the implications and meanings in a game very important (especially their verification and validation and checking their compliance to standards, rules and regulations). For example, an Educational Game should not harm the educational process of a student and it should really improve his/her educational skills. Or a Game with a Purpose should provide only that purpose. Or Serious Games – like any other serious matter – should be made consciously and they should be responsible towards their related standards, rule and regulations. Cognitive Games beside improving skill, can harm cognitive, mental or even physical health of player, therefor they should be studied from this perspective. Verification and validation of the meanings and compliance checking is only achievable through rigorous and formal meanings; and this fact highlights the importance of accessing rigorous and formal semantics for games.

2. **Literature Review**

In this section, we took a brief literature review in two aspects. First, we review using rigorous methods, rigorous and formal semantics in game and game developing. Then, we review the literature in the field of supplying the health of video games in Iran and the world briefly.

**2.1. Literature about Using Formal and Rigorous Methods in Game and Game Developing**

Using formal and rigorous methods (i.e. using rigorous and formal semantics for games' representations and specifications) in game developing are as follows:

- Use formal and rigorous methods to achieve auto adjust for difficulty of game levels [6]
- Increase the ability to use pre-existing components to develop new games by defined, formal and rigorous communication between game logic components (which different for each game) and components such as graphic motor, and user interface libraries and compilers (which are reusable in different games) [7]
- Use rigorous and formal methods to increase understanding of game logic for developers [7]
- Use rigorous and formal semantics to provide the use of an existing implemented gaming platform (which grammar and compiler is not available) besides the available compilers and make a combined compiler by mixing "an implemented platform" and "several compilers" [8].
- A rigorous specification of a game mechanisms to use them in gamification [9,10].
- Use direct Executable Formal Specifications in designing, validation and evolution of the games [11], [31].
- Use formal representation to combine and integrate learning process and gaming process in Educational Games [12] and to check and verify it [27].
- Use formal representation and modeling for designing Cognitive Games (to use cognitive patterns and toys for making Cognitive Games, we should be able to combine and integrate cognitive process with gaming process; one way to achieve this is using formal models for integration.) [13], [14], [15], [18]
- Use games' formal specification languages to produce games quickly and automatically [16]
- Use games' formal specification to reach ideas, decide, solve and calculate the related issues, and using them in benchmarking processes in Artificial Intelligence, specifically as a benchmark for Artificial General Intelligence (AGI) algorithms [16]

- Use formal specification to find a proper game for serious and educational purposes and validate the game adaptation with those purposes [17], [27].
- Use formal specification to formally represent the nature of rules and regulations of a game and rely on code generators to make codes which run the game. In this way, you can make sure that the video players (artificial intelligence agents) observe the rules of the game, and also we can have video players (intelligent agents) who have more complex and advanced behaviors (since a complete set of rules for the game is officially available for them). [19]
- Use formal methods, models and specifications in the path to promote intelligence and artificial intelligence in video games [20].
- Use formal methods and mathematics to improve and promote game experience for the player and even supply the adaptation of this experience with the player. This matter is achieved through formal and rigorous mathematical-based analysis of the game and the players [21], [26], [29], [30].
- Combine user-generated content with pre-existing game scenarios to produce active Educational Games in a customized way and in accordance with needs and contents of the user [22]. For this matter, we need specified and calculated models from the content and scenario of the game.
- Use formal methods to implement and improve approaches based on Model Driven Development (MDD) in game industry [23].
- Use formal models and rigorous and formal specifications to calculate various types of qualitative analysis; check and prove the correctness of properties for a game especially, soundness for game scenario [24].
- Documenting the design of Serious Games to reach correctness from design [25]
- Use formal specifications to study games and their studies at high level of abstraction and regardless of implementation [5], [28].

Figure 1 shows an example of a formal specification for a game which was provided in reference [5]. Of course, the semantic model which was used in this sample (which is based on set theory) is just one of many models available in this field.

Heno [5] has used formal specifications to review the games and their designs at high level of abstraction and regardless of implementation. He stated that using formal specification for games allow us to achieve several goals, such as: automating many levels in game designing (such as adjusting the difficulty of game levels, automatic generation of code based on high level specifications); having a media which is understandable, public and proper for transmitting game design ideas; achieving evolution and preservation for a game through applying complexity management and separation of concerns on structural features of game software (which are stated as specification features in formal specification). Moreover, since the parts and abstractions of a game are defined as features of formal specification, we can achieve reusability of a game assets. Figure 1 depicts a sample of a formal specification of a game which has been used by Heno [5].

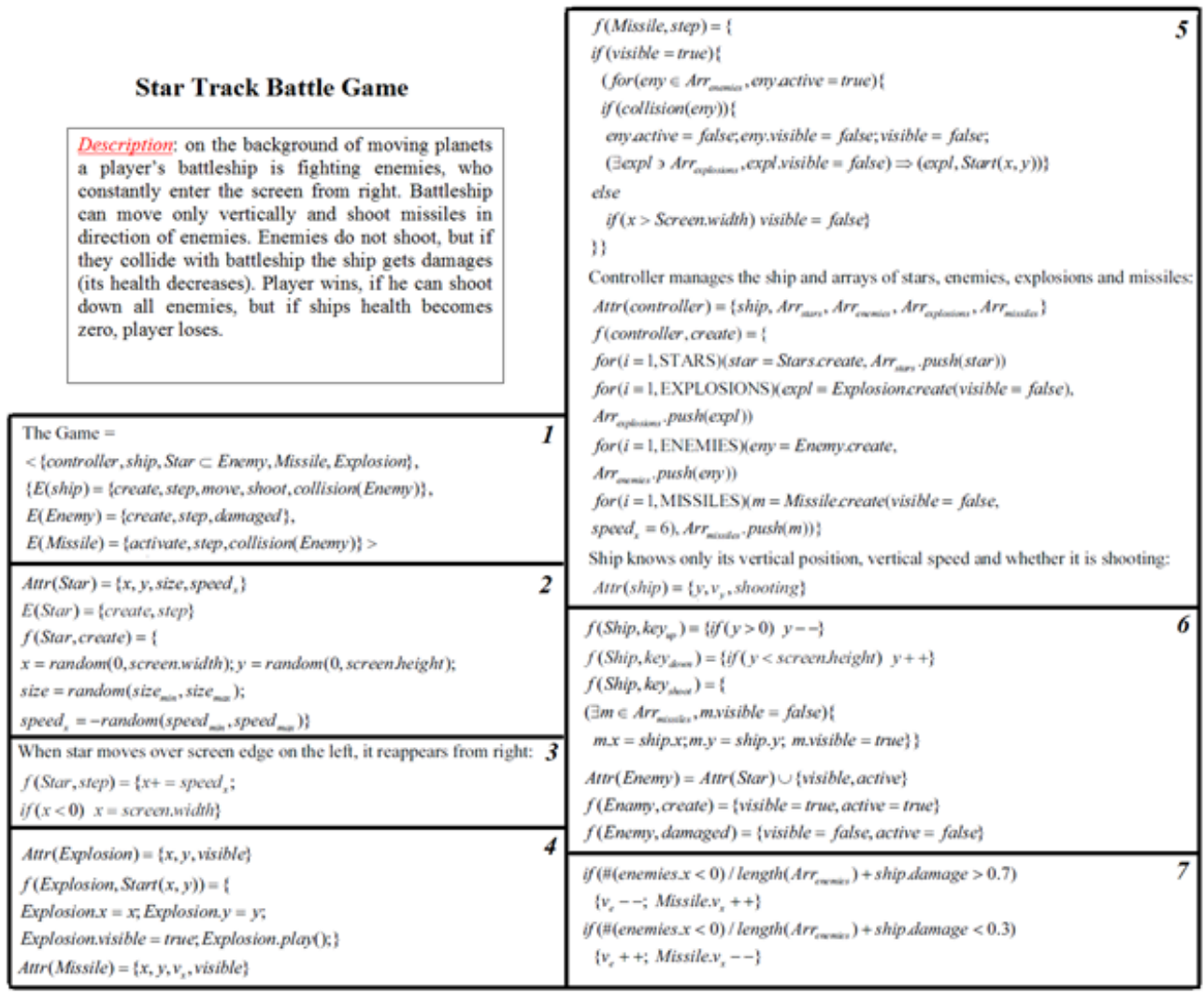

**Figure (1): Brief Description and Formal Specification of a Game; Source: [5]**

## 2.2. Literature about Supplying the Health of Video Games

Many countries use video game content rating systems, these systems sometimes are based on self-declaration of the company or game developer (such as AppStore Ratings) and sometimes are based on specialists' reviews (Such as ESRA in Iran, USK in Germany, PEGI in European Union, MEKO In Finland, ESRB in USA, Mexico and Canada and GRB in South Korea). These ratings in most countries play informing role for the parents and help self-regulation in game industry to achieve health in games. Moreover, there are some experiences in Iran and some other countries about passing some rules to support and supply the health of games. Unfortunately in today world, profits of unhealthy games in the one hand, and weakness of cultural and religious institutions and marginalization of ethics and spirituality on the other hand, create serious obstacle for proper legislation in this field. Our country is exempt from this defect and there exist no obstacle in achieving the goal of supplying the health of the games.

### 3. Proposed Organizing for Assessing and Supplying the Health of Video Games

Figure 2 depicts the general outlook of our proposed organizing. Since health of a game has various aspects, using specialists and experts from different fields is a necessity. This matter helps us to achieve a set of rules and regulations for health of a game. Therefore, a council which includes specialized commissions is needed to regulate the health of a game, this council is called "regulatory council".

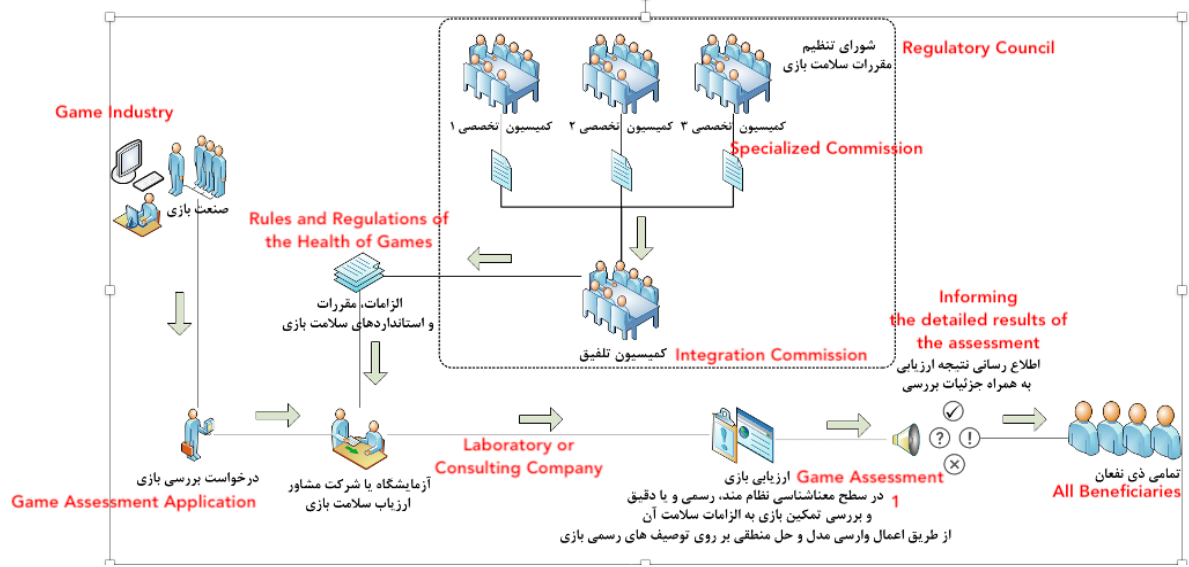

**Figure 2. The Proposed organization for evaluating and ensuring the health of computer games**

Using the industry delegates and their ideas in specialized commissions is necessary; lack of attention to this matter can endanger the whole process of regulation. Moreover, conducting community though surveys and considering the interests of players is another important factor which should be provided through polling organizations and researchers for this council. Of course, various rules and regulations should be matched and integrated together. Therefore, rules and regulations which are provided by specialized commissions will be sent to integration commission. In this commission which includes the delegates of all other commissions and other beneficiaries, the final decision is made and an integrated and compatible set of rules and regulations for the health of games will be prepared.

Private sector (specifically, laboratories and consulting companies that assess health of games) are the agents of assessing games and the council only set the rules and regulations. Game developing companies should present their games to any laboratory or consulting company and apply for an assessment.

The laboratory or consulting company with the presence of delegates from the developing company, provide formal and rigorous specifications and systematic models for the product. Then, the game will be assessed formally by using these formal specifications and the formal specifications of health of a game rules and regulations, and implementing methods of semantic solution for compliance (such as using checking compliance tools, model checking tools and logical solution tools). If the game contradicts with any rules and regulations, it is exactly identified that which part of the game contradicts with which part of rules and regulations; and this contradiction can be proved logically. If there is no contradiction between the game and the rules, this compliance is also provable. Details available in the results and systematic process of this review allow the game developing company to be able to repeat the assessment and correct the errors quickly; in an integrated agile cooperation, game health confirmation is issued and it is informed to all beneficiaries. This method through reviewing meanings, implications and impacts rigorously and formally can provide detailed guidelines for each games and this will in turn help the parents to manage the negative aspect of each game. (For example, for a game with extreme excitement, it can be mentioned that playing for more than 1 hour can make the child stressful or anxious.)

### 3.1. The Advantages of the Proposed Organizing

The following facts are the advantages of this proposed organizing (the collection these facts supply various concerns of different beneficiaries):

1. Agility: It is important for industry and game developers. Timing for reviewing the game should not be too long, since it will deprive the industry from competition by decreasing innovation and "Time-to-Market". Besides, it will result in the unwanted depreciation of the invested capital for the game. To achieve agility, extracting formal specifications and compliance check of the game with rules and regulations is assigned to laboratories or consulting companies. Private sector is both specialized and scalable and it can grow according to the needs of game industry. Moreover, private sector benefits the agility of this reviewing process. The existence of multiple assessment agents in this field, allow game industry to implement B2B communication and facilitate the process automatically. Besides, the competition which emerges among private assessment companies increase assessment efficiency and it also allow government, observing authorities and regulatory authorities to replace those companies who violate the laws. In this way, the health of the process of assessment is guaranteed by enforcing laws and competition.

2. Flexibility: Separating the concern of operating from regulating help the regulatory council to decide based on the conditions, and be flexible by accepting change and updating rules. This concern separation also help the assessment agencies in private sector to be able to work in their field freely and adapt themselves to conditions (for example, take over each other in assessing health of the games by using new technologies and with proper financial management become a fix character in this market.)

3. Scalability: The structure of the council consists of separate specialized commissions which is scalable. The emerging of a new concern can add a new commission. Also, consulting companies in this field change according to the needs of game industry; there exists a self-regulatory mechanism among market, game industry and game assessment consulting companies.
4. Repeatability of reviews: Reviews are computerized calculations and can be retrieved. Formal modelings are also based on defined principles which are retrievable. This fact is very helpful in solving legal claims – which are inevitable in this big market and industry.
5. Provability of the Reviews: Formal semantic methods (mathematical methods, logical and logical solution, model check and so on) are all parts of argumentative and inferential sciences and techniques. So in all of these methods, the result of the review is always based on a proof and this proof is provided along the result.
6. Transparency: Rules and regulations which are confirmed by the regulatory council are informed to all beneficiaries; and everyone can criticize, correct or program these rules. Moreover, the assessment companies should accept the responsibility of all their provided services and activities.
7. Adaptation: The council is live entity which accepts inputs, and it can create a decision making ring with society with the help of specialized arms (such as polling organizations and researchers). In this way, it can adapt itself with current conditions and facts. Moreover, game industry can participate in decision making by attending the council effectively. In this way, this industry can reach its goal and claim its rights.
8. Available details for reviews
9. Assessing various layers and implicit and explicit implications of system of the game: Game is a complicated system and has various meanings. The implications of a game can have positive or negative impacts on audience. Using semantics and computational methods allows us review and assess different layers of meaning wisely. In this way, tacit semantics and tacit knowledge of a game are assessed.
10. Avoiding subjectivity or individual skills and rely on rules and regulations: This is true that violating this matter does not harm game assessments but relying on rules and regulations helps us to provide higher level of service in this field.
11. Beneficiaries' ability to plan because of its transparency: Game industry, decision makers, society, media, families and even users (especially in Serious Games) can be informed of the procedures relating the assessment of a game and its health; in this way, they can plan based on these information for their future activities. This fact decrease the risk of working in game industry significantly, and provide power for industry besides guaranteeing the health of games. In this way, dangerous games will be omitted from this industry. This matter is the direct result of regulations and transparency in the system.

12. Employment for specialists (by meeting the needs of the society): Assessing and supplying the health of a game is a very important need for each society. The consulting assessors in this field (which should be from humanities and technical sciences) serve the society and at the same time achieve a proper and stable job. This employment will active the investment which the society has put through its educational system to educate these people.

4. **The Technical Form of Assessing the Health of a Game and Refining a Solution for it**

Video games which are developed in game industry should observe some rules like any other products. These rules are for the quality and health of the games. So in game industry like other industries we face an issue called regulations and regulatory. Wherever regulations and regulatory are existed, the issue of compliance and solving this issue is also existed [32]. Game regulations set some rules for qualified and healthy games. Each video game can be considered as a logical and computational system and the rules for it are also the compliance requirements for this system, then we can provide the formal specification for these two logical systems. Then we can check the compliance of these two systems by using model checking methods and logical solution. Refinement of the solution is depicted in Figure 3.

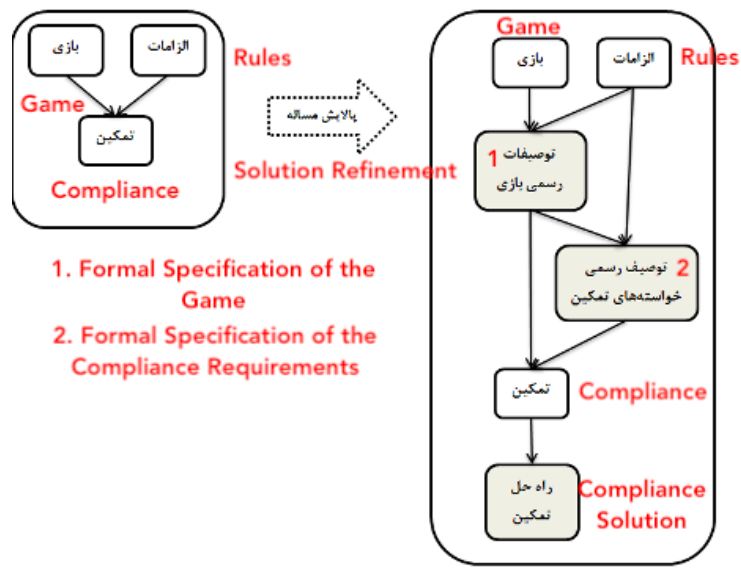

**Figure (3): The Technical Form of Assessing the Health of a Game and Refining a Solution for it**

## 5. An Example

Assume that the following rules are considered for a game; these rules have been set for example by industrial regulations, national standards or a company product procedure. (These rules are just some examples and they are not complete):

1. The excitement of the game should not be extreme and continuous. Of course extreme excitement [but not very extreme] is allowed in some parts of the game.
2. The violence of the game should be monotonous.
3. The violence of the game should not be a lot.
4. The game should have the minimum attraction in all parts of it.
5. The attraction of the game should be at its highest level at the start and at the end of the game.
6. There should be problem solving and operating in the game, but we should be able to pass the game by avoiding one of them.
7. The more easier understanding the game atmosphere, the more desirable the game is (observing Attention Economy).
8. The game should not be confusing.
9. Hot elements of the game (those elements which attracts most attention) should not preach violence or be anti-value.
10. There should be a balance between variety in game elements and its simplicity.
11. The game reacts to the player's achievements, and this reaction is beneficial and effective in the game procedure.
12. The achievements do not increase greedy emotions in the player.
13. The game should not be hard or easy. Achieving most of the achievements should be possible for most of the players. (challenging and at same time not frustrating)
14. The first achievement is possible in the first 1 minute for almost all players. It should has challenge too.

The compliance requirements should be extracted from these rules in the form of a formal specification. Various semantic models should be applied to represent, review and solve these compliance requirements. We can use assessment techniques and patterns (such as GQM) or requirements engineering (such as KAOS) to refine the rules, in this way we can achieve formal specifications of compliance requirements; the supply of those requirements can supply the observation of the rules. Table 2 shows which semantic models and logical-mathematical tools can be used for each of the above mentioned rules. These consist of tools, models and formal languages of system modelings such as Alloy, SAN and Reo.

**Table 2- Which formal model or tool could be used to describe and examine each of the 14 obedience demands given in the example?**

|  | Alloy\ Reo\ SAN | Cellular Automata | Fuzzy Logic | Deontic Logic | Prioritization Logics | Cognitive Concepts Models | Differential calculus | Temporal Logics\ Temporal Models | Algebraization |
|---|---|---|---|---|---|---|---|---|---|
| 1 | | | ❊ | ❊ | ❊ | ❊ | ❊ | ❊ | ❊ |
| 2 | | | ❊ | ❊ | ❊ | ❊ | ❊ | ❊ | ❊ |
| 3 | | | ❊ | ❊ | ❊ | ❊ | ❊ | ❊ | ❊ |
| 4 | | | ❊ | ❊ | ❊ | ❊ | ❊ | ❊ | ❊ |
| 5 | | | ❊ | ❊ | ❊ | ❊ | ❊ | ❊ | ❊ |
| 6 | Reo | ❊ | ❊ | ❊ | | | | | |
| 7 | Alloy | | ❊ | ❊ | ❊ | ❊ | | | ❊ |
| 8 | Reo\Alloy | ❊ | ❊ | ❊ | | ❊ | ❊ | ❊ | ❊ |
| 9 | Reo\Alloy | | ❊ | ❊ | ❊ | ❊ | ❊ | ❊ | ❊ |
| 10 | Alloy | | ❊ | ❊ | ❊ | | | | |
| 11 | Reo | | ❊ | ❊ | | ❊ | | | |
| 12 | Reo\SAN | | ❊ | | | ❊ | ❊ | ❊ | ❊ |
| 13 | SAN | | ❊ | ❊ | ❊ | ❊ | | ❊ | ❊ |
| 14 | SAN | | ❊ | ❊ | | ❊ | | ❊ | |

## 6. Conclusion

The proposed organizing for assessing and supplying the health of video games can be a progressive step for supplying the concerns of all of the beneficiaries in this field. Relying on formal methods and specifically formal specifications and models, model checking techniques and solving systems' compliance can be a new chapter in the field of assessing the health of video games. This approach can go further than the past (which were mostly based on video games content rating systems and systematic assessment and not rigorous and formal assessments) and lead us to a healthier virtual world. Agility, flexibility, scalability, repeatability of reviews, transparency, adaptation, available details for reviews, assessing various layers and implicit and explicit implications of system of the game, avoiding subjectivity or individual skills, relying on rules and regulations, ability to plan for beneficiaries because of its transparency and employment for specialists are among the features of this organizing.